# Structural and magneto-transport studies of iron intercalated $Bi_2Se_3$ single crystals

Shailja Sharma and C.S. Yadav*

*School of Basic Sciences, Indian Institute of Technology Mandi, Mandi-175005 (H.P.) India*

*Email: shekhar@iitmandi.ac.in*

A detailed investigation on the structural and magneto-transport properties of iron intercalated $Bi_2Se_3$ single crystals have been presented. The x-ray diffraction and Raman studies confirm the intercalation of Fe in the van der Waals gaps between the layers. The electrical resistivity of the compounds decreases upon intercalation, and Hall resistivity shows the enhancement of the charge carriers upon intercalation. The magnetoresistance shows the non-saturating linear behavior at higher magnetic field and low temperature. Intercalation of Fe increases the onset of the linear magnetoresistance behavior, indicating the reduction in quantum effects. The Kohler scaling employed on the magnetoresistance data indicates single scattering process for all these compounds in the measured temperature range of 3- 300 K.

**Introduction:** Topological insulators are the quantum materials with the insulating bulk and conducting surface states.[1,2] The peculiar property that distinguished these from trivial insulators is that the bulk band gap is inverted due to strong spin-orbit coupling and the gapless surface states are protected topologically by time reversal symmetry (TRS).[2] Three dimensional topological insulators have surface states with odd number of Dirac cones in which spin and momentum are locked in a chiral structure.[3] The spin-momentum locking prohibits the backscattering of electrons since it requires spin-flip.[4] Bismuth selenide ($Bi_2Se_3$) is one of the most studied compounds among the topological insulators family.[5] $Bi_2Se_3$ has a large bulk band gap (300 meV) and simple surface states possessing single Dirac cone.[6] Topological surface states with a single Dirac cone have been observed through angle resolved photoemission spectroscopy (ARPES), scanning tunneling microscopy (STM), confirming the theoretical prediction by Kane and Mele.[6-9] Besides these techniques, surface states in TIs have been studied using the quantum transport behavior also. Some of the signatures of two-dimensional surface states discussed in literature are Shubnikov-de Haas (SdH) oscillations, weak antilocalization (WAL), non-saturation linear magnetoresistance (LMR) and universal conductance fluctuation (UCF).[10-13] The quantum transport is mainly dominated by the bulk transport in such systems as the Fermi level is usually found to be in its bulk conduction band due to selenium vacancies. These vacancies are believed to give rise to electron doping that makes the crystals to grow in a n-type material. To make the Fermi level shift inside the band gap, Ca, Mg, Cd are used to turn it into p-type $Bi_2Se_3$.[14-16]

In recent years, intercalated $Bi_2Se_3$ have got rekindled attention among the researchers due to observation of unconventional superconductivity with intercalation of Cu, Sr, or Nb in $Bi_2Se_3$.[17-19] Further, the doping by magnetic element in the van der Waals (*vdW*) gaps finds interest in engineering the band properties of the material. One ARPES study on the magnetically doped $Bi_2Se_3$ showed the opening of an energy gap at Dirac point resulting from the TRS breaking by magnetic doping.[20] There are several experimental reports on the magnetically doped topological insulators that results in bulk magnetism: $Bi_{2-x}M_xSe_3$ (M = Mn, Fe, Cr, V).[21-27] Despite the enough information on bulk magnetism on Fe doped $Bi_2Se_3$, only few have discussed magneto-transport in such systems.[28,29] Earlier reports on Fe-doped $Bi_2Se_3$ have studied the bulk ferromagnetism in conformity with the anomalous Hall effect, metal-insulator transition with suppression of magnetoresistance (MR) values at high Fe content.[30,31] However, the magneto-transport properties in Fe-intercalated $Bi_2Se_3$ systems remain largely unexplored.

In this work, Fe intercalated $Bi_2Se_3$ has been comprehensively studied by means of x-ray diffraction (XRD), Raman spectroscopy, and low-temperature magneto-transport. It is confirmed that the lattice structure does not disrupt/change with intercalation up to highest concentration (x = 0.15). The resistivity and Hall measurements data respectively discussed the metallicity throughout the temperature range 3-300 K and the increase in carrier concentration with the Fe content in $Bi_2Se_3$ leads to the fact that Fe atoms have preferably occupy the interstitial positions in the *vdW* gaps. The effect of magnetic doping on linear magnetoresistance phenomenon up to 10 tesla magnetic field has been discussed. Moreover, it is observed that single scattering rate is followed across the fermi surface in these systems as Kohler's scaling is obeyed.

**Experimental details:** Single crystals of $Fe_xBi_2Se_3$ (x = 0, 0.10, 0.15) were synthesized in three-step process using melt-growth method. The first step was to prepare $Bi_2Se_3$ for all the required compositions. High purity selenium (≥ 99.999%), bismuth (≥ 99.99%) in the form of pellets/granules were weighed according to their nominal compositions and sealed in evacuated (>$10^{-5}$ mbar) quartz tubes. An excess amount (~3%) of Se has been used in order to compensate for the Se vacancies. The compounds were heated at 850 $^0$C for 48 h, followed by cooling to 550 $^0$C at the rate of 3 $^0$C/h, where they were kept for 72 hours. Next, the compounds were furnace-off cooled to room temperature. Secondly, the grown $Bi_2Se_3$ and iron (≥ 99.98%) were melted at 850 $^0$C for 120 h, slow cooled (3

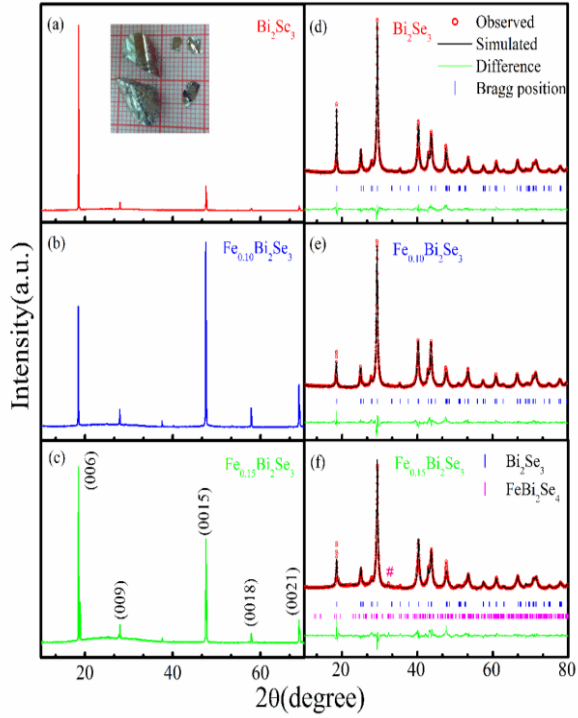

Fig. 1: X-ray diffraction pattern of single crystals (a, b, c) and Rietveld refined powder XRD (d, e, f) of $Fe_xBi_2Se_3$(x = 0, 0.10, 0.15), respectively. Inset shows the image of as grown crystals.

°C/h) to 550 °C and left for 24 h and then furnace off cooled to room temperature. To ensure homogeneity, samples were re-grinded, sealed and kept for third heat treatment, where they were kept at 850 °C for 24 h and slow cooled (3 °C/h) to 300 °C, then furnace off cooled to room temperature. The obtained crystals exhibited metallic appearance and were cleaved easily.

Phase purity and crystal structure analysis were carried out by powder x-ray diffraction (XRD) using a Rigaku Smart lab diffractometer with Cu-K$\alpha$ radiation ($\lambda$ =1.5418 Å) at room temperature. Raman analysis was carried out employing Horiba HR-Evolution spectrometer using 532 nm solid state laser. The linear four probe technique was used to study the resistivity measurements in the temperature range 3-300 $K$ and field range 0-10 $Tesla$ in a Quantum Design built Physical Properties Measurement System (PPMS). The Magnetoresistance (MR) and the Hall-effect measurements were carried out at different temperatures in the PPMS.

**Results and discussion:** The orientation and crystallinity of single crystals are shown in Figure 1(a,b,c), which clearly reveals (0 0 $l$) as the most preferred direction. The peaks are labelled with (0 0 $3n$) miller indices. The Rietveld refinement for the powder XRD patterns was performed using *Fullprof suite* as depicted in figure 1(d,e,f). The Bi$_2$Se$_3$ crystallizes into the rhombohedral structure with the space group ($R\bar{3}m$).[32] It contains five atoms in the primitive unit cell.

TABLE 1. Lattice parameters of the $Fe_xBi_2Se_3$ (x=0, 0.10, 0.15) obtained from Rietveld refinement fit to XRD patterns

| $Fe_xBi_2Se_3$ | a [Å] | c [Å] | c/a |
|---|---|---|---|
| x = 0 | 4.1375(6) | 28.6307(9) | 6.919 |
| x = 0.10 | 4.1404(8) | 28.6560(8) | 6.921 |
| x = 0.15 | 4.1427(9) | 28.6358(3) | 6.912 |

TABLE 2. Characteristic peak positions of different modes in Raman spectra of the $Fe_xBi_2Se_3$ (x = 0, 0.10, 0.15)

| $Fe_xBi_2Se_3$ | $A^1_{1g}[cm^{-1}]$ | $E^2_g[cm^{-1}]$ | $A^2_{1g}[cm^{-1}]$ |
|---|---|---|---|
| x = 0 | 72.03 | 131.75 | 174.50 |
| x = 0.10 | 71.72 | 131.49 | 173.31 |
| x = 0.15 | 71.52 | 130.93 | 173.33 |

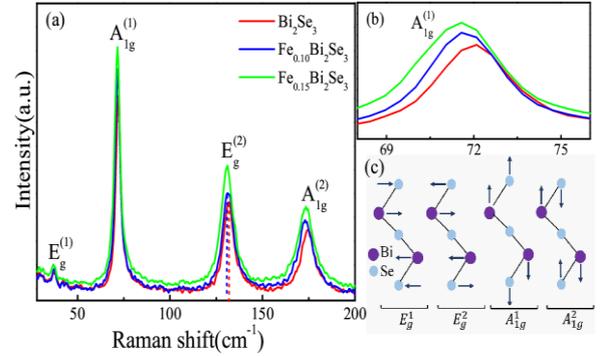

Fig. 2: (a) Raman spectra of $Fe_xBi_2Se_3$ (x=0, 0.10, 0.15) single crystals. (b) Shift in peaks for mode $A^1_{1g}$ (c) Schematics of all four Raman active modes.

This rhombohedral structure is formed by the stacking of quintuple layers (QL) along the c-axis perpendicular to the ab plane. Each unit cell of Bi$_2$Se$_3$ is comprised of three QLs, each in the sequence *Se$^{(1)}$-Bi-Se$^{(2)}$-Bi-Se$^{(1)}$*, linked by the *vdW* forces. The chemical nature of both Se atoms are different, *Se$^{(2)}$* is more ionic compared to *Se$^{(1)}$* which is bonded to *Se$^{(1)}$* in next QL via weak *vdW* bonds. The XRD pattern for Bi$_2$Se$_3$ and Fe$_{0.10}$Bi$_2$Se$_3$ confirms the phase purity, however Fe$_{0.15}$Bi$_2$Se$_3$ shows minor secondary peak which can be identified to ~1% of the secondary monoclinic phase (*C2/m*) corresponding to FeBi$_2$Se$_4$.[33] Crystallographic parameters obtained from the refinement are presented in table 1. The lattice parameters *a* and *c* increase considerably with Fe content. This supports the possibility that Fe intercalates in the *vdW* gaps between the QLs. The ionic radii of Fe$^{2+}$ (0.92 Å) and Fe$^{3+}$ (0.78 Å) are much smaller than the Bi$^{3+}$ (1.17 Å) and Se$^{2-}$ (1.98 Å), if Fe is substituted at ionic (either Bi or Se) site, the lattice constants *a* and *c* should be reduced, however, there was no reduction in lattice constants. Further, there is no pronounced change in the axial ratio (c/a) with the doping of Fe atoms.[34]

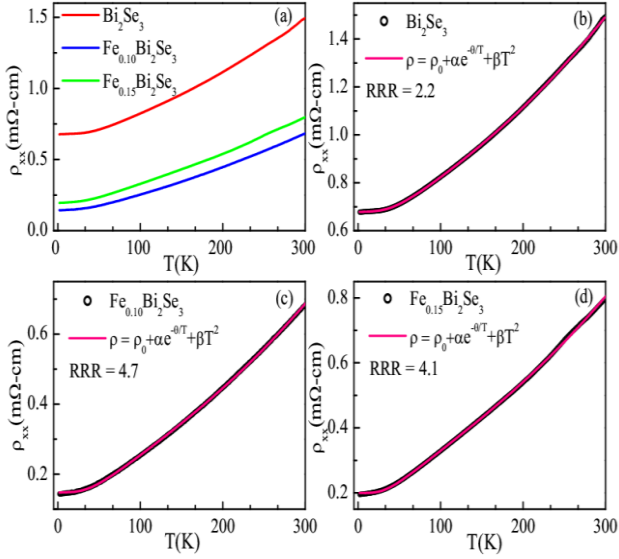
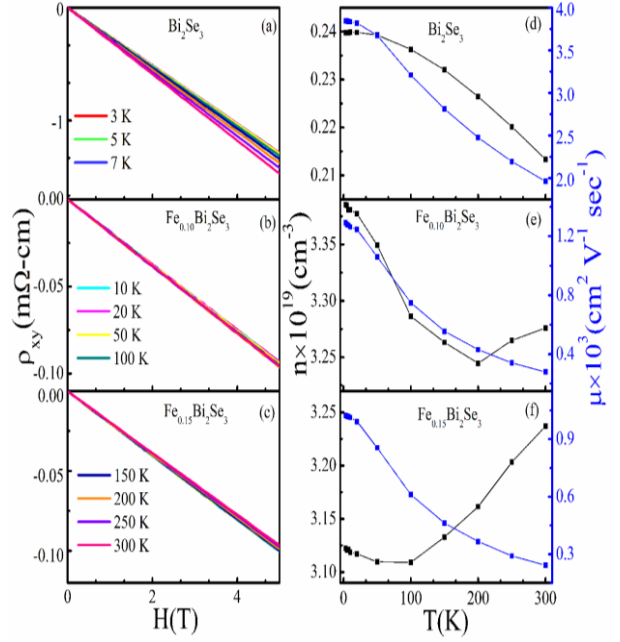

Fig. 3: (a) Temperature-dependent longitudinal resistivity of $Fe_xBi_2Se_3$ (x= 0, 0.10, 0.15) (b), (c), (d) show the Bloch-Grüneisen fits in the temperature range between 2 K and room temperature.

Fig. 4: The Hall resistivity (a, b, c) and the variation of carrier concentration n (left axis) and mobility μ (right axis) with temperature (d, e, f) for $Fe_xBi_2Se_3$ (x= 0, 0.10, 0.15), respectively.

Raman spectra shown in Figure 2, provides information about the lattice vibrational modes. Four intrinsic active phonon modes $E_g^1$, $A_{1g}^1$, $E_g^2$ and $A_{1g}^2$, at 37.2 cm$^{-1}$, 66.6 cm$^{-1}$, 131.5 cm$^{-1}$, 174.5 cm$^{-1}$, respectively have been observed in accordance with the previous studies.[35-37] The $E_g$ and $A_{1g}$ modes corresponds to atomic vibrations along-the-plane and perpendicular to the layers, respectively. Generally, the bulk phonon modes ($E_g$ and $A_{1g}$) are expected to shift to lower frequencies upon intercalation, and towards higher frequencies upon substitution.[38, 39] As shown in the figure 2(b), three modes are shifted to lower frequency. The red shift implies that Fe atoms occupy the sites between QLs and weakens the interaction between layers which results in the decrease of phonon mode energy. The high quality of the crystal can be observed clearly as the low frequency $E_g^1$ mode has been observed in present work which is absent in most of the reports.[38, 40] Table 2 shows the shift in peak positions for the different observed modes. Since $E_g^1$ mode is weak, the frequency shift for this mode is not shown. Thus, the shift of peaks to lower wavenumber supports that Fe atoms are intercalated in the $Bi_2Se_3$ lattice.

The temperature dependent longitudinal resistivity $\rho(T)$ of the compounds measured in the range 2-300 K are shown in figure 3(a). The compounds show the metallic behavior throughout the temperature range, as indicated by $d\rho/dT > 0$, which points toward the presence of Se vacancies. The electrical resistivity value of Fe-intercalated $Bi_2Se_3$ is lower than the pristine $Bi_2Se_3$. The samples have residual resistivity value (RRR) ($\rho(300 K)/\rho(2 K)$) in the range 2.2 - 4.7; the lowest one for the parent compound. The temperature dependence of resistivity can be fitted to a simplified model developed for bulk crystals with, $\rho_{xx} = \rho_0 + \alpha\exp(-\theta/T) + \beta T^2$, where $\rho_0$ is the residual resistivity arising from impurity scattering.[41] The exponential and quadratic terms arises as a result of electron-phonon scattering and electron-electron scattering respectively. The fitting parameter $\theta$ corresponds to Debye phonon frequency, $\omega = k_B\theta/T$. A comparison of fitting parameters for different compounds are tabulated in table 3. Additionally, no upturn was found down to lowest temperature as has been reported earlier in Fe substituted $Bi_2Se_3$.[31, 42] The electrical resistivity for undoped and doped samples show metallic behavior which agrees well with the carrier density obtained from Hall measurements. This kind of behavior is commonly observed in crystals with high carrier concentration (>10$^{17}$cm$^{-3}$) where bulk band conductivity dominates.

Figure 4 shows the magnetic field dependence of Hall resistivity at different temperatures. Hall resistivity was found to be linear in the magnetic field. The Hall resistivity have been anti-symmetrized using the relation, $\rho_{xy}^{Hall} = (\rho_{xy}(+H) - \rho_{xy}(-H))/2$, in order to eliminate the offset voltage due to misalignment.[11] The slope of the Hall coefficient, $R_H = \frac{\rho_{xy}}{H}$, remains negative throughout the temperature range which confirms electron dominated charge transport. The Hall carrier density is calculated using relation, $R_H = 1/ne$. The value of carrier density increases by one order of magnitude with Fe intercalation, as plotted in figure 4(d, e, f). The order of carrier density variation from ~10$^{19}$ cm$^{-3}$ to 10$^{17}$ cm$^{-3}$ for $Bi_2Se_3$ is in well agreement with reports.[43] It is important to mention here that the order of carrier density for $Bi_2Se_3$ decides the metallic behavior of electrical resistivity, since for crystals with carrier density less than 10$^{17}$ cm$^{-3}$, it turns to insulating behavior at low temperatures.[43, 44] The low temperature mobilities of the compounds were found to be in the range $\mu_e$~10$^2$-10$^3$ cm$^2$V$^{-1}$s$^{-1}$, highest for the pure compound. Since, $Bi_2Se_3$ has lower RRR value but one order of higher magnitude of mobility as compared to Fe intercalated compounds, the variation in the electrical resistivity of the compounds is more influenced by the carrier concentration than the internal defect or scattering. There is difference of one order of magnitude in carrier concentration between pristine and Fe intercalated $Bi_2Se_3$, the carrier concentrations show very small variation with temperature. Figure 4 shows the plot for carrier mobility ($\mu$) and concentration $n = 1/R_H e$ obtained from the zero-field longitudinal resistivity $\rho_{xx}(T)$ and the measured transverse resistivity $\rho_{xy}(B)$, respectively. Temperature dependence of carrier density is negligible below 30 K, and almost saturates, coinciding with low-temperature resistivity behavior, suggesting a common origin. The obtained carrier concentration is in the range of (0.21-0.24) $\times 10^{19}$ cm$^{-3}$ for $Bi_2Se_3$, (3.27 - 3.48)$\times 10^{19}$ cm$^{-3}$ for $Fe_{0.10}Bi_2Se_3$, (3.12 - 3.23)$\times 10^{19}$ cm$^{-3}$ for $Fe_{0.15}Bi_2Se_3$ in Transverse magnetoresistance (MR) was measured at different temperatures with the applied magnetic field (0-10$T$) perpendicular to the direction of current. MR is expressed as the change in resistivity under applied field ($\Delta\rho(H)$) normalized by zero field resistivity ($\rho(H = 0)$) or $(\rho(H) - \rho(H = 0))/\rho(H = 0)$. The magnetoresistance values have been symmetrized using relation $\rho_{xx}^{MR} = (\rho_{xx}(+H) + \rho_{xx}(-H))/2$ in order to eliminate any Hall contribution in resistivity values.[11] The MR exhibits positive values at all temperatures, which increases on lowering the temperature (Figure 5). The MR shows the quadratic field dependence at low fields (shown as dotted black lines) and non-saturating linear behavior at high fields (shown as solid black lines). As observed from the plots, the MR increases on increasing Fe concentration although the qualitative behavior remains same. This the measured temperature range (3 - 300 K). The weak temperature variation of carrier concentration can also be seen from the overall behavior of the Hall resistivity as it remains almost same, with subtle or slight change for 3 K and 300 K measurements. The decrease in mobility on increasing temperature correlates with the increases of phonon scattering as the temperature increases and in agreement with the behavior of resistivity curves. It is known that Fe exists in two states as $Fe^{2+}$ and $Fe^{3+}$. Here, Fe atoms may intercalate into van der Waals gap as $Fe^{2+}$ as donor or it may substitute Bi as $Fe^{3+}$ and act as acceptor, causing slight change in the carrier density. However, we have observed an order of enhancement in carrier density. Also, we have shown that Fe atoms intercalate in vdW gaps by Raman spectroscopy. Intercalation of Fe atoms tend to induce significant charge transfer between Fe and $Bi_2Se_3$, which is clearly reflected in Hall measurements.

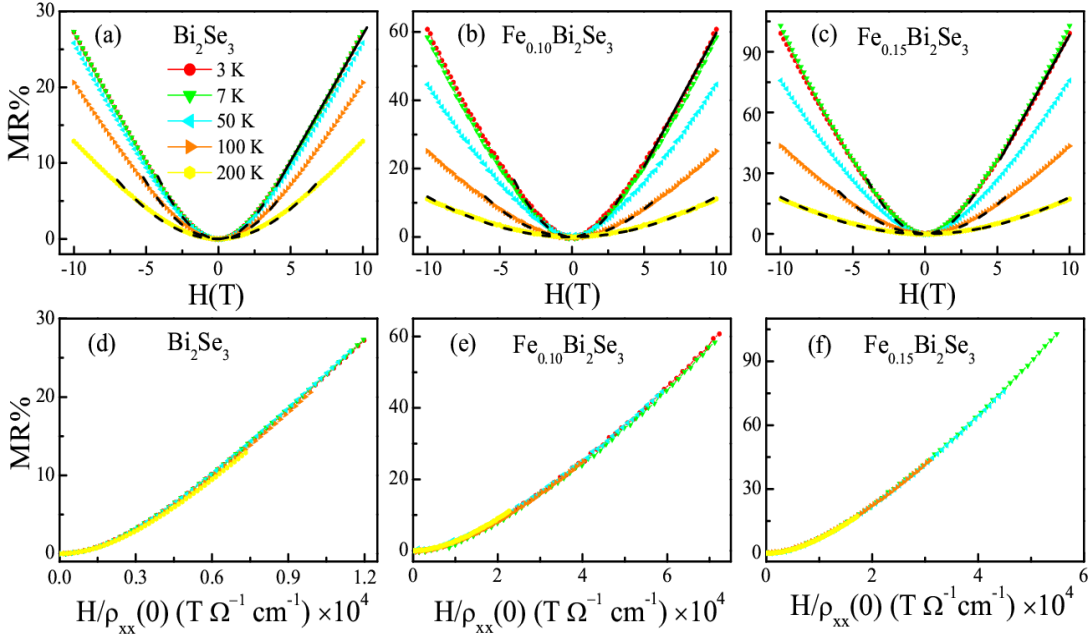

Fig.5: Magnetoresistance (MR) as a function of magnetic field at different temperatures for (a) $Bi_2Se_3$, (b) $Fe_{0.10}Bi_2Se_3$, (c) $Fe_{0.15}Bi_2Se_3$. The black dashed lines show the quadratic field dependence at various temperatures up to different fields. Solid black lines show the linear fit to the high field data. Kohler plot (d, e, f) of all the measured magnetoresistance.

agrees with increase in carrier concentration with doping. The quadratic field dependence of MR is understood using semi-classical model that attributed to the deflection of conduction electrons by the Lorentz force under the applied magnetic fields. This quadratic behavior saturates at high fields, followed by the linear magnetoresistance. The linear MR may be described based on classical model by Parish and Littlewood, which explains the linearity due to the disorder induce mobility fluctuations in an inhomogeneous sample.[45] Since, these samples are single crystalline in nature, we do not see any correlation of mobility fluctuations to explain the linear magnetoresistance with the classical model. A plausible explanation for the observed linear MR in these systems could be the quantum origin, although no SdH oscillations are observed up to 10 T. This linear MR is often explained using the Quantum model of Abrikosov. In this model, linear MR is explained in the quantum limit when all the carriers occupy the lowest Landau level. Thus, $\rho_{xx} \propto \frac{NH}{n^2}$ provided $n \ll (eH/c\hbar)^{3/2}$ and $T \ll eH\hbar/m^*$, where $N$ and $n$ are the density of scattering centers and electrons respectively, and $H$ is the applied magnetic field.[46] This model relates the linear MR to the linear energy dispersion relation of the gapless topological surface states.[47] Abrikosov proposed that in the quantum limit when magnetic fields are so large such that Landau level are well formed, the carrier concentration should be small enough so that electrons occupy only the lowest Landau level. This implies that this model is applicable in the extreme quantum limit. Moreover, it is interesting to note that not all the systems with LMR have observed the SdH oscillations.[48, 49] The observed carrier density is of the order $10^{19}$ cm$^{-3}$, and indicate that our these systems are far away from the quantum limit such that electrons could hardly occupy the lowest Landau level (observed up to 10 T magnetic fields). However, the presence of linear magnetoresistance at higher field points the importance of surface states as well. Theoretically, as per Abrikosov's model conduction from gapless surface states gives rise to linear magnetoresistance. It is quite possible that these quantum effects are gaining sufficiently strength at higher magnetic fields and reflected in our data. Furthermore, the weak antilocalisation (WAL) effect have been observed in such systems.[50] It is observed predominantly in lower dimensional systems, where bulk carrier density is low implying less contribution from bulk. Figure 6 shows the temperature variation for the magnetic field ($B^*$) for the onset of linear MR. Our results show that $B^*$ increases as the temperature increases, and with the increase in Fe concentration also. Since there is increase in carrier concentration on Fe doping $B^*$ has larger value compared to the pristine $Bi_2Se_3$.

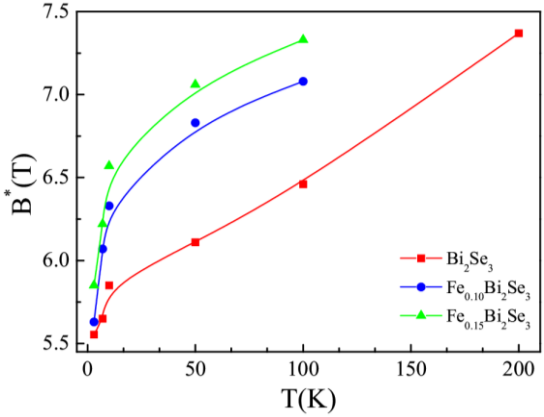

Fig.6: Temperature variation of crossover magnetic field of linear MR for $Fe_xBi_2Se_3$ (x = 0, 0.05, 0.10)

To further study the type of scattering process in our system, MR data is analyzed using Kohler's scaling of data at different temperatures. The change in isothermal resistivity, $\Delta\rho(H)/\rho(H=0)$, in an applied field (H) depends upon the quantity $\omega_c\tau$ which is the product of $\omega_c \propto H$ and $\tau(T) \propto 1/\rho(T)$ resulting in $\frac{\Delta\rho(H)}{\rho(H=0)} = f(\frac{H}{\rho(H=0)})$, $\omega_c$ is the cyclotron frequency, at which magnetic field causes the charge carriers to sweep across the Fermi surface and $\tau$ is the relaxation time.[51] Figure 5 shows the Kohler plots for all the measured magnetoresistances. Kohler's rule is satisfied if there is a single scattering rate ($\tau$) at all point on the Fermi surface. All MR curves at different temperatures collapse onto a single curve suggesting single scattering rate in these systems. Although the intercalation of Fe gives rise to increase in the carrier concentration of $Bi_2Se_3$, the magneto-transport processes remain largely unchanged.

**Conclusion:**

In summary, structural and magneto-transport properties of $Fe_xBi_2Se_3$ (x = 0, 0.10, 0.15) single crystals synthesized using melt-grown technique have been reported. The x-ray diffraction studies confirmed the rhombohedral crystal structure of $Fe_xBi_2Se_3$. Rietveld refinement shows the lattice expansion which suggests that Fe atoms are intercalated between *vdW* gaps. The phonon properties of the single crystals investigated through Raman spectroscopy confirms intercalation of Fe atoms at *vdW* gaps. By analyzing the resistivity and MR measurements, it can be found that Fe content tends to increase metallicity in $Bi_2Se_3$, and magnetoresistance value also increases. Thus, present studies show bulk conductance is dominant over the surface conductance. High *n*-type carrier concentrations, $10^{18}$-$10^{19}$cm$^{-3}$ were obtained from Hall coefficient measurements.

**Acknowledgements**
We acknowledge Advanced Material Research Center (AMRC), IIT Mandi for the experimental facilities. CSY acknowledges the DST-SERB project YSS/2015/000814 for the financial support. SS acknowledgesIIT Mandi for the HTRA fellowship.